# Abrupt global events in the Earth's history: a physics perspective

Gregory Ryskin

Robert R. McCormick School of Engineering and Applied Science

Northwestern University, Evanston, Illinois 60208

ABSTRACT

The timeline of the Earth's history reveals quasi-periodicity of the geological record over the last 542 Myr, on timescales close, in the order of magnitude, to 1 Myr. What is the origin of this quasi-periodicity? What is the nature of the global events that define the boundaries of the geological time scale? I propose that a single mechanism is responsible for all three types of such events: mass extinctions, geomagnetic polarity reversals, and sea-level fluctuations. The mechanism is fast, and involves a significant energy release. The mechanism is unlikely to have astronomical causes, both because of the energies involved, and because it acts quasi-periodically. It must then be sought within the Earth itself. And it must be capable of reversing the Earth's magnetic field. The last requirement makes it incompatible with the consensus model of the origin of the geomagnetic field – the hydromagnetic dynamo operating in the Earth's fluid core. In the second part of the paper, I show that a vast amount of seemingly unconnected geophysical and geological data can be understood in a unified way if the source of the Earth's main magnetic field is a ~200-km-thick lithosphere, repeatedly magnetized as a result of methane-driven oceanic eruptions, which produce ocean flow capable of dynamo action. The eruptions are driven by the interplay of buoyancy forces and exsolution of dissolved gas, which accumulates in the oceanic water masses prone to stagnation and anoxia. Polarity reversals, mass extinctions, and sequence boundaries are consequences of these eruptions. Unlike the consensus model of geomagnetism, this scenario is consistent with the paleomagnetic data showing that "directional changes during a [geomagnetic polarity] reversal can be astonishingly fast, possibly occurring as a nearly instantaneous jump from one inclined dipolar state to another in the opposite hemisphere".

# 1. Framing the questions
## 1.1 Introduction

What makes physics different? Steven Weinberg put it well: "One of the primary goals of physics is to understand the wonderful variety of nature in a unified way" (Weinberg 1999). By contrast, historical sciences such as biology or geology focus on the particular, and deal with an overwhelming amount of detail, most of it contingent on the actual path of development (the history) of their subject matter. An attempt to find a unifying theme in a maze of historical detail may encounter strong resistance. But on those rare occasions when such an attempt succeeds, the result is a transformation of revolutionary proportions. Examples are the molecular biology revolution, and the plate tectonics revolution in Earth science.

With few exceptions, the Earth science community was firmly opposed to Alfred Wegener's proposal of continental drift for 50 years, until in 1963 Lawrence Morley, and independently Vine and Matthews (1963), combined the sea-floor-spreading hypothesis of Hess (1962) with the geomagnetic polarity reversals (whose reality was denied for even longer time). The Vine-Matthews-Morley hypothesis started the plate tectonics revolution; the conversion of the Earth science community was complete in a few years (Hallam 1989, Oreskes 2001). Prior to these developments, Wegener's proposal was deemed unacceptable because "It was too large, too unifying, too ambitious. Features that were later viewed as virtues of plate tectonics were attacked as flaws of continental drift." (Oreskes 2001, p. 11).

Some of the processes that were initially inferred on the basis of geological record are now directly measurable. For example, the theory of plate tectonics implied that continents were moving with velocities of the order of a few centimeters per year; such movements can now be tracked using the global positioning system. Nevertheless, the most interesting questions in Earth science, and the answers to them, must be sought in the geological record. As in cosmology (another historical science), that record is unique, and the system is not subject to experimentation. In the case of cosmology, the assumptions of symmetry (homogeneity and isotropy) on the large scale reduce the complexity of the problem enormously; for geology, nothing comparable is possible. It is hardly surprising that, with the exception of plate tectonics, no progress has been made toward a "unified theory" of geology.

## 1.2 Quasi-periodicity of the geological record

One feature of the Earth's history may present an opportunity for theoretical analysis. The geological time scale (Gradstein et al. 2004; see Fig. 1) reveals quasi-periodicity of the geological record over the last 542 Myr, on timescales close, in the order of magnitude, to 1 Myr. For example, the intervals between geomagnetic polarity reversals typically range from 0.2 to 2 Myr; between geological stage boundaries – from 1 to 10 Myr, etc. Exceptions do exist, e.g., there were no polarity reversals between 84 and 124 Myr ago.

The geological time scale is the final result of assimilation and interpretation of a staggering amount of geological and geophysical data. This is done within the conceptual framework of *stratigraphy*, the study of sedimentary rock strata, their temporal correlation and ordering (ICS 2009). The very existence of stratigraphy is predicated on the presence in the geological record of clearly identifiable markers, reflecting some global events. Given the imperfection of the geological record, it is obvious that abrupt, geologically instantaneous global events would leave the best possible markers. But it is far from obvious why the Earth should produce such events. The fact that it has, at least for the last 542 Myr, is remarkable; it is telling us something very important, but in order to understand the message, the right questions must be asked first. My purpose in the present paper is to frame the questions (Part 1), and also to describe my own attempts to provide answers (Part 2).

Among the most important branches of stratigraphy are biostratigraphy (based on the fossil content of the rock), magnetostraigraphy (based on the geomagnetic polarity recorded in the rock), and sequence stratigraphy (based on sequences of strata deposited on continental margins by the cycles of sea-level rise and fall). Biostratigraphy was developed in the first half of the nineteenth century (Hancock 1977, Hallam 1989), magnetostratigraphy – in 1960's (Glen 1982), and sequence stratigraphy – in 1970's. All three have been spectacularly successful in practical terms. Biostratigraphy, in particular, provides relative ages of the strata; after absolute ages of a number of tie points were determined using the radiometric dating, it became possible to establish the geological time scale for the last 542 Myr (the Phanerozoic eon). The rocks older than 542 Myr contain hardly any fossils, and biostratigraphy cannot be used.

Yet, there is still no understanding of what caused the global events on which these powerful methodologies are based. This is not unusual: astronomy was put to practical uses such as the calendar and navigation long before terrestrial and celestial mechanics were unified by Isaac Newton. Some attempts at explanation have been made in the past. Georges Cuvier suggested that discontinuities in the fossil record reflected mass extinctions produced by environmental catastrophes, such as inundations by the sea; he did not discuss possible causes of the catastrophes *per se*. His influential essay of 1812 was entitled "*Discours sur les révolutions de la surface du globe*"; an English translation came out the following year (Cuvier 1813) and went through several editions. At that time, *catastrophism* had eloquent supporters in Britain, but they were soon outnumbered by the *uniformitarians*, whose motto was "the present is the key to the past", and who took this to mean that only the processes observable now may have operated in the past (Hallam 1989). Charles Darwin, in particular, denied the reality of mass extinctions altogether, and ascribed any evidence for them to gaps in the geological record (Raup 1994). This is not surprising: even though Darwin used the fossil succession as evidence of evolution, the theory of evolution offers no explanation for mass extinctions (Raup 1994).

**1.3 Mass extinctions and biostratigraphy**

In most of the geological literature, the designation "mass extinction" is reserved for the most severe extinctions in the Earth's history, such as the Permian-Triassic ca. 251 Myr ago, the Cretaceous-Tertiary (Cretaceous-Paleogene) ca. 65 Myr ago, and a few others. Each of these extinctions eliminated ~70 to 90% of the total number of species; consequently, these extinctions define the most significant boundaries of the geological time scale. In particular, the two extinctions mentioned above define the boundaries between the geological eras, Paleozoic, Mesozoic, and Cenozoic, which together comprise the Phanerozoic eon. However, as emphasized by Raup (1994), other extinctions – e.g., the ones marking stage boundaries – were not qualitatively different from the "Big Five", and should be included in the same category. Hallam and Wignall (1997, p. 1) define mass extinction as "an extinction of a significant proportion of the world's biota in a geologically insignificant period of time". Generally speaking, a biostratigraphic boundary is marked by a mass extinction:

> "Five percent [of the total number of species] is roughly the extinction level that normally defines [the boundary of] the 'biostratigraphic zone' – the smallest unit in geologic time recognizable by fossils on a global or near-global basis. In many parts of the geologic column, paleontologists have estimated the average duration of a stratigraphic zone to be about one million years." (Raup 1991, p. 170)

Stage boundaries are defined by mass extinctions of greater severity (typically, a few tens of percent), followed, in the rough order of severity, by the boundaries between epochs, periods, and eras. (The time units "age", "epoch", and "period" correspond to the stratigraphic units "stage", "series", and "system".)

It may seem that the extinction level of 5% is relatively low, and does not qualify as a mass extinction when compared to extinctions of 70%. This view is common in the geological literature, but is incorrect (Raup 1991, 1994). The percentage of species lost is not the right statistic to consider when trying to weigh the severity of the environmental perturbation (an abrupt global event) that caused mass mortality in the biota, and resulted in the disappearance of a number of species. Instead, one should compare the mortality statistics at the level of individuals;

this paints a completely different picture. It is, of course, the individuals that are actually affected; on the timescale of the extinction-causing event, the species is an abstract entity.

The following analysis is based on the statistical approach originated by Raup (1979; 1991, pp. 70-75); the final result (1) appears to be new. Assume for simplicity's sake that the extinction of a species requires the death of all its individual members (organisms). Let the total number of organisms in a species be $N$. If killing is completely random, the probability of extinction is $P_{ext} = (1 - P_{si})^N$, where $P_{si}$ is the probability of survival for an individual. Hence $\ln(1 - P_{si}) = N^{-1} \ln P_{ext}$; here $N$ is a large number, while $\ln(1 - P_{si})$ is small because $P_{si} \ll 1$. Then $\ln(1 - P_{si}) = -P_{si}$, and therefore $P_{si} = -N^{-1} \ln P_{ext}$. Consider now the expected number of surviving individuals in a species, $n_{si} = NP_{si}$. This number is very simply related to $P_{ext}$, and the relation does not include $N$, viz., $n_{si} = -\ln P_{ext}$. Or, equivalently,

$$P_{ext} = e^{-n_{si}} \qquad (1)$$

In an extinction where 5% of species are lost, only about 3 individuals in a species are expected to survive, on average. The expected number of survivors drops to 1 for the extinction level of 37%, as could occur, e.g., at a stage or a series boundary. It drops to 0.1 for the extinction level of 90%, possibly reached in the greatest extinction in the Earth's history, the Permian-Triassic boundary.

On the other hand, a catastrophic event slightly less fatal, such that 10 individuals in a species were expected to survive, would leave no evidence of extinction in the fossil record: given the imperfection of the latter, the extinction level of 0.005% is unobservable.

These examples may be counter-intuitive, but they express a simple truth: for a species with a large number of members, to become extinct is highly improbable, as survival of even a single member is enough to prevent that from happening. And only species extinction is relevant for the fossil record; mortality at the level of individuals is not. A catastrophic event may cause nearly complete mortality in the biota, but still fail to cause mass extinction; such an event would leave hardly any trace in the fossil record.

The above simple analysis cannot be true in detail; it assumes random killing and ignores the fact that some of the species are more resilient than others (Raup 1979; 1991, pp. 70-75). But its main message is valid: a sudden environmental event capable of causing even a "minor" mass extinction must be utterly catastrophic at the level of individuals. This message is essentially contained, even if never stated explicitly, in the works of Raup (1991, 1994).

I conclude: The occurrence, over the last 542 Myr, of hundreds of mass extinctions at the biostratigraphic boundaries, suggests the existence of some unknown mechanism, acting quasi-periodically on timescales of order 1 Myr, and causing some terrifying global catastrophes.

**1.4 Magnetostratigraphy and sequence stratigraphy**

Similarly to mass extinctions, the reality of geomagnetic polarity reversals was denied for nearly 60 years, until the Vine–Matthews–Morley hypothesis connected the polarity reversals with the sea-floor spreading, and ushered in the plate tectonics revolution (Hallam 1989, Oreskes 2001). Magnetostratigraphy has since become an indispensable tool of stratigraphic correlation, even though the mechanism of reversals remains as mysterious as ever. I discuss in Part 2 why the "standard model" of geomagnetism – the hydromagnetic dynamo operating in the Earth's fluid core, and occasionally self-reversing – is unlikely to be true.

As far as sequence stratigraphy is concerned, the denial stage is not over yet, though this has not prevented sequence stratigraphy from becoming a tool of choice in oil and gas exploration. Here by denial I mean the refusal to accept the main tenet of sequence stratigraphy, viz., that depositional sequences reflect cycles of sea-level rise and fall (Vail et al., 1977a, 1977b). Indeed, some geologists maintain that the central ideas of sequence stratigraphy are nothing but a myth (Dickinson 2003). The conceptual problem is real: global sea-level changes ~100 m on timescales ~1 Myr cannot be explained in the absence of continental ice sheets. Thus Dewey and Pitman (1998) write:

> "We can discern no mechanism that can cause the necessary short-wavelength [~1 Myr], large-amplitude sea-level changes implicit in globally synchronous eustatic third-order cycles" of sequence stratigraphy.

Miller et al. (2004) concur:

> "Either continental ice sheets paced sea-level changes during the Late Cretaceous [an ice-free epoch by all other indications], or our understanding of causal mechanisms for global sea-level change is fundamentally flawed."

The sequence stratigraphy community has ignored this objection (Miall and Miall 2001), and the paradox remains unresolved.

They have not been able, however, to ignore another objection: Their original sea-level curve, based directly on the seismic record, showed the sea-level falls (but not rises) as geologically instantaneous (Vail et al. 1977a, 1977b). This caused such a storm of criticism that in subsequent publications they invoked factors invisible in the seismic record, such as tectonic subsidence, and produced a "more sinusoidal" sea-level curve (Jervey 1988; Haq et al. 1987;

Hallam 1992, p. 25). (Uniformitarianism is alive and well – any suggestion of abrupt change is strongly resisted. Cf. Weart 2003.) The original saw-tooth sea-level curve was not entirely abandoned, however; it survives under the name "coastal onlap" (Haq et al. 1987), or simply "onlap" (Haq and Schutter 2008); see Fig. 2.

The pattern is becoming clear now: Every new branch of stratigraphy utilizes a new type of marker in the geological record. The reality of the abrupt global events that left these markers is invariably denied by the majority of experts; this, however, does not stop the practitioners of the new stratigraphy from achieving great practical success. As a result, the voices of denial become progressively weaker, and may disappear entirely; still no explanation is forthcoming of the nature of the events.

**1.5 A single mechanism?**

The phenomena that lie behind the practice of stratigraphy should be of great interest to physicists, who are not constrained by the "dangerous doctrine of uniformitarianism" (Ager 1993b, p. xvi). In addition, I propose that a single mechanism is responsible for all three types of events – those underlying biostratigraphy, magnetostratigraphy, and sequence stratigraphy. This proposal is based on the following empirical evidence. (For brevity, the word "mass" will be omitted from now on, "extinction" always meaning "mass extinction".)

The "strong correlation between extinctions and magnetic reversals" is well known to geologists (Ager 1993a, p. 37), but remains unexplained. Equally well known and unexplained is the correlation between extinctions and sequence boundaries.

> "In the seven cores studied the magnetic reversals and faunal boundaries are consistently related to each other … The coincidence or near coincidence of faunal changes with reversals in these cores suggests a causal relation." (Opdyke et al. 1966)
>
> "Many geologists expect sequence boundaries to correspond with system, series, and stage boundaries and … zonal subdivisions …" (Gradstein et al. 2004, p. 236)
>
> "The most often recognized surface is the combined sequence boundary and subsequent flooding (transgressive) surface. Most standard stage type sections located in passive-margin settings, have a transgressive surface as their lower boundary." (Hardenbol et al. 1998, p. 4)
>
> "The regional and the global stage boundaries are sequence boundaries that reflect the global event of the sea-level fall." (Vakarcs et al. 1998)
>
> "The initiation of each major transgressive episode coincides with a major mass extinction …" (Gradstein et al. 2004, p. 288)

The geological record, in fact, contains multiple instances of perfect coincidence. Here are a few examples, listing the dates, in million years ago, of a polarity reversal, a stage boundary, and a sequence boundary: 28.45, 28.45, 28.45; 23.80, 23.80, 23.80; 20.52, 20.52, 20.52; 14.80, 14.80, 14.80 (Gradstein et al. 2004, pp. 69-71; de Graciansky et al. 1998, Chart 2). The Oligocene-Miocene boundary, dated 23.03 Myr ago, coincides with a polarity reversal (Gradstein et al. 2004, p. 424). The Permian-Triassic boundary coincides with a polarity reversal (Ward et al. 2005). The four most recent stage boundaries, dated 3.60, 2.59, 1.81, and 0.78 Myr ago, all coincide with polarity reversals (Gradstein et al. 2004, pp. 28-29; ICS 2009); the last three – also with sequence boundaries. The list could be continued.

The above is strong evidence that all three types of events are caused by the same mechanism. The alternative – three separate mechanisms, which often act at exactly the same time – is so improbable that it hardly deserves attention.

It is also true that not every stage boundary will coincide with a polarity reversal and/or a sequence boundary. The reasons are simple: all three types of events being caused by the same mechanism does not mean that all three must occur, and leave a clear record, every time the mechanism acts. For example, a slight variation in the mortality produced by the event may cause it to leave no trace in the fossil record (see above). Polarity reversal need not occur each time the mechanism acts (see below). And some of the events may have left a poor record yet to be discovered.

**1.6 The problem of accurate dating**

An additional problem is the difficulty of mapping the sedimentary column to the time axis. Radiometric dating has achieved spectacular successes in geology, but sedimentary rocks cannot be dated in this way. Only igneous rocks or volcanic ashes can be dated radiometrically because the radioactive isotope and its decay products must be "locked in" within the crystalline matrix; otherwise their ratios will be distorted by diffusion and other effects (Bourgeois 1990). Radiometric dating in stratigraphy is typically limited to volcanic ash horizons that bracket biostratigraphically-calibrated sedimentary sections (Gradstein et al. 2004, p. 89). If one uses the assumption of constant sedimentation rate between such horizons, a catastrophic, geologically-instantaneous deposit will be interpreted as representing thousands or even millions of years (Ager 1993a, Ager 1993b, Bourgeois 1990). Then the ages assigned, say, to a biostratigraphic boundary and a polarity reversal may fail to coincide, even though the extinction and the reversal occurred simultaneously.

Currently, no reliable technique exists for mapping the sedimentary column, between the radiometrically-dated tie points, to the time axis. A case in point is provided by the strata associated with the Cretaceous-Tertiary boundary – probably, the most thoroughly studied segment of the geological record. In several geographical locations, a particular sedimentary complex within these strata is interpreted by some geologists as deposited over 0.3 Myr, by others – over a few hours or days (by a mega-tsunami). The debate on this issue has been going on for over two decades, and shows no sign of abating, with world-class experts in both camps marshalling the ever increasing amounts of data to support their respective interpretations. (For the latest salvos from each side of the debate, see Keller et al. 2009a, 2009b and Schulte et al. 2010.) The implied sedimentation rates differ by eight orders of magnitude.

**1.7 Playing with time**
One immediate conclusion is this: with the exception of the radiometrically-dated tie points, the mapping $t(z)$ of the sedimentary column to the time axis must be viewed as an unknown function, which is monotonically increasing but otherwise of the most general nature. This function possesses multiple jump discontinuities, corresponding to the periods of time when either no deposition occurred, or the deposited strata were subsequently removed (eroded). The sedimentary record is "more gaps than record … a lot of holes tied together with sediment" (Ager 1993a, Ch. 3). The discontinuities of $t(z)$ – the periods of "lost time" – are marked in the sedimentary record by *unconformities* (Fig.3) Unconformities are of paramount importance in interpretation of the sedimentary record; in particular, they define sequence boundaries in sequence stratigraphy.

While $t(z)$ is piecewise continuous, the value of its derivative may range over many orders of magnitude, and there is no reason to expect $t(z)$, or its inverse, to possess any degree of smoothness. As noted above, between the radiometrically-dated tie points $t(z)$ is essentially unknown. It is impossible to use a function like that for any practical purposes (e.g., construction of graphs), so one often assigns to $t(z)$ maximal smoothness – linearity – between the tie points (which is equivalent to assuming the constant sedimentation rate). This is only a consequence of the currently unavoidable ignorance, and should not preclude contemplation of phenomena that would lead to highly non-smooth $t(z)$.

Consider, for example, the sea-level curve of sequence stratigraphy. In its construction, of both the original saw-tooth version and the "more sinusoidal" version, constant sedimentation rates were assumed. As discussed above, there is no rational basis for this assumption; if it is dropped, the time axis can be locally stretched or compressed at will, with the sea-level curve being deformed as a result. Imagine now the sea-level curve transformed into a comb-like series of instantaneous peaks, coinciding in time with the abrupt sea-level falls of the original saw-tooth ("onlap") curve (Haq et al. 1987, Haq and Schutter 2008, see Fig. 2). In other words, could an entire cycle of sea-level rise and fall be geologically instantaneous?

It could not, of course – if the sea level is understood in its usual sense of a global, quasi-static datum. But things change dramatically if fluid dynamics is brought into the picture. As pointed out by Dott (1996), and also by Ager (1993a, 1993b), the sequences of sequence stratigraphy could be deposited by mega-tsunamis. Such a scenario immediately resolves the great controversy over the sea-level changes in the absence of continental ice sheets (see above). And it is fully supported by the field data: The strata directly overlying a sequence boundary typically contain pebble conglomerates, lag gravels, and rip-up clasts of the underlying lithologies (Baum and Vail 1988), which all signify a fast, high-energy flow. Such a flow occurs when coastal areas are flooded by a tsunami (Bourgeois 2009), not during a quasi-static sea-level rise ~100 m in a million years.

The mega-tsunami scenario also solves another long-standing puzzle, the presence of erratics (boulders, etc.), normally interpreted as glacial dropstones, in deposits from the ice-free epochs:

> "Concentrations of erratics and [fossilized] wood appear to occur in distinct horizons or boulder beds. These coincide with the basal portions of transgressive units." (Markwick and Rowley 1998)

> "Allochthonous logs … occur in extraordinary abundance as sedimentary components in transgressive marine shelf deposits …" (Savrda 1991)

**1.8 Summary of Part 1**

There are strong indications that a single mechanism is responsible for the mass extinctions, geomagnetic polarity reversals, and mega-tsunamis that underlie biostratigraphy, magnetostratigraphy, and sequence stratigraphy. This mechanism has been acting quasi-periodically over at least 542 Myr, on timescales close, in the order of magnitude, to 1 Myr. The mechanism is fast, and involves a significant energy release. It is unlikely to have astronomical causes, both because of the energies involved, and because it acts quasi-periodically. It must then be sought within the Earth itself.

This already looks like a problem of great interest, but the intrinsic interest is not the only motivation here: Unless we understand the mechanism, we shall have no chance of preventing it from acting again. And it may act again, strictly speaking, any moment: the last four of its actions are dated 3.60, 2.59, 1.81 and 0.78 Myr ago. Still, on the margins of a million-year timescale, there is probably some time left.

## 2. A common origin for the Earth's magnetic field and stratigraphic boundaries
### 2.1 Geomagnetism as a problem of physics

The origin of the Earth's magnetic field is one of the oldest problems of physics. In 1269, Petrus Peregrinus (Pierre de Maricourt) wrote *Epistola de Magnete* (Peregrinus 1269), "the first scientific treatise describing observations and experiments carried out for the purpose of clarifying natural phenomenon. The conclusions were derived logically based on observations and experiments." (Kono 2007). Written 400 years before scientific journals were invented, *Epistola de Magnete* had a form of a letter to a friend, and spread via manuscript copies. Peregrinus discovered (and named) the poles of a magnet, as well as the impossibility of separating them, i.e., the nonexistence of magnetic charges. In order to explain the propensity of the magnetic needle to align along the meridian, he proposed that "it is from the poles of the heavens that the poles of the magnet receive their virtue". (In the geocentric system, the celestial sphere rotates about the axis passing through the celestial poles.) He further proposed that the magnet as a whole is influenced by "the whole heavens" and, if properly oriented on frictionless pivots, would "move according to the motion of the heavens" – in essence, the first hint at Mach's principle (Peregrinus 1269).

Peregrinus's explanation of the magnetic needle behavior remained a reasonable hypothesis until the discovery of declination in the 15$^{th}$ century; the first recorded measurements of declination were made by Christopher Columbus, who also discovered the dependence of declination on the geographic location (Kono 2007). In 1600, William Gilbert in *De Magnete*, the first scientific monograph, proposed that the Earth itself is a "great [permanent] magnet" (Gilbert 1600, pp. 211-212). Gilbert's hypothesis suffered a major setback when secular variation of the Earth's main magnetic field was discovered by Henry Gellibrand (1635). It was ultimately abandoned because (Chapman and Bartels 1940, Ch. 21):

(a) secular variation could not be explained;
(b) the required magnetization of the lithosphere appeared to be too high (since temperature increases with depth, only the outer shell of the Earth can be permanently magnetized);
(c) "it would be hard to explain how the magnetization could be everywhere so nearly parallel to the magnetic axis, unless some powerful hypothetical process was assumed by which the Earth was magnetized at some past epoch, although no trace of this process is now left".

It would be even harder to explain the polarity reversals, but their reality was largely denied at the time.

In the early decades of the 20$^{th}$ century, the origin of the Earth's magnetic field was considered one of the most important problems of physics (Einstein 1924). Some far-reaching proposals were made but only one survived to this day, by Joseph Larmor (1919), who suggested the following origin for magnetic fields of celestial bodies: Electric currents are induced in a conducting fluid moving through a magnetic field; these currents give rise to secondary magnetic fields, which add to the original field. If amplification of the field is faster than its resistive decay, the flow may act as a generator of magnetic field – a hydromagnetic dynamo. The initial field can be arbitrarily small; the field growth is exponential until its back action on the flow (via the Lorentz force) becomes significant.

With regard to the Earth in particular, Larmor (1919) singled out secular variation as "the very extraordinary feature of the earth's magnetic field", and suggested that the above mechanism would account for secular variation "merely by change of internal conducting channels; though, on the other hand, it would require fluidity and residual circulation in deep-seated regions".

Larmor did not develop this conjecture in any detail; more than 25 years passed before it was applied to the Earth's core, by Frenkel (1945), and independently by Elsasser (1946). At the turn of the century, hydromagnetic dynamo action was finally demonstrated in the laboratory (Stefani et al. 2008), though neither the flow nor the magnetic field in the experiments bear much resemblance to their counterparts in celestial bodies.

Currently, the consensus is that the geomagnetic main field is produced by the hydromagnetic dynamo in the Earth's fluid outer core (Roberts and Glatzmaier 2000). Secular variation is used to estimate the characteristic large-scale velocity in the outer core, and to conclude that dynamo action is possible. This model has not been seriously questioned for decades, even though it makes no testable predictions. (With one exception: the characteristic timescale of magnetic diffusion in the core, ~10 kyr, imposes an approximate lower limit on the duration of a polarity reversal.)

In Ryskin (2009), I proposed a different mechanism of secular variation: "ocean water being a conductor of electricity, the magnetic field induced by the ocean as it flows through the Earth's main field may depend on time and manifest itself globally as secular variation". This proposal was supported by calculation of secular variation using the equations of magnetohydrodynamics, and by analysis of observational data. "If secular variation is caused by the ocean flow, the entire concept of the dynamo operating in the Earth's core is called into question: there exists no other evidence of hydrodynamic flow in the core" (Ryskin 2009).

Note that it is impossible to determine the location of the source by observing the field at the Earth's surface. For example, the external field of a uniformly magnetized spherical shell is exactly that of a point dipole, and depends only on the product of magnetization and volume; the radii of the shell cannot be inferred. In addition, there exists an infinite variety of magnetizations of a spherical shell that produce no external field at all (Runcorn 1975). The commonly assumed separation of the observed field, on the basis of its spherical harmonic representation, into the field produced in the Earth's core and the crustal field, has no theoretical basis.

Below I show that a vast amount of seemingly unconnected geophysical and geological data can be understood in a unified way if the source of the Earth's main magnetic field is a ~200-km-thick lithosphere, repeatedly magnetized as a result of methane-driven oceanic eruptions (Ryskin 2003), which produce ocean flow capable of dynamo action. Polarity reversals, extinctions, and sequence boundaries are consequences of these eruptions. Unlike the consensus model, this scenario is consistent with the paleomagnetic data showing that

> "directional changes during a reversal can be astonishingly fast, possibly occurring as a nearly instantaneous jump from one inclined dipolar state to another in the opposite hemisphere" (Acton et al. 2000).

**2.2 The magnetizable lithosphere**

At the atmospheric pressure, the Curie temperature of iron oxides and other iron-containing minerals is not higher than 675°C, but it rises with pressure; rates of increase ~23 K per GPa have been measured (Schult 1970). This means that at a depth of 200 km, the Curie temperature of iron oxides can be ~800°C. Metal alloys that form in the deep lithosphere may have the Curie temperature as high as 1,100°C (Haggerty 1978). The actual temperature variation with depth is not well constrained. Temperatures inferred from seismic data are ~800 to 1000°C at 200 km depth (Goes et al. 2005; Kuskov et al. 2006). Observations suggest that the upper mantle is, in fact, magnetic (Pochtarev et al. 1997; Blakely et al. 2005). Thus, a deep Curie isotherm is not ruled out by the data; let us take for its depth 200 km.

The volume of the 200-km-thick outer shell of the Earth is ~$10^{20}$ m$^3$. The dipole moment of the Earth's magnetic field is ~$0.8 \times 10^{23}$ Am$^2$. If the source of the field is the 200-km-thick lithosphere, the average magnetization of this lithosphere must be ~1 kA/m. This is a very large value, much larger than the average magnetization of the near-surface rock. However, under the conditions of high temperature and pressure prevalent in the deep lithosphere, iron-oxide minerals may acquire remanent magnetizations ~$10^2$ to $10^3$ kA/m (Robinson et al. 2002; Gilder and LeGoff 2008), provided the magnetizing field is much stronger than the geomagnetic field. Since iron oxides make up a few percent of the lithosphere, the average magnetization of the lithosphere ~1 kA/m is feasible, if only barely. Just barely accounting for the observed intensity of the Earth's magnetic field is, however, not a flaw but a virtue: the important question of why the field has this particular intensity is then resolved.

Thus, objection (b) is not fatal to Gilbert's hypothesis. But what could magnetize the lithosphere, and do so repeatedly, reversing the field polarity?

**2.3 Methane-driven oceanic eruption as dynamo**

In Ryskin (2003), I proposed the existence of methane-driven oceanic eruptions, a quasi-periodic Earth-based phenomenon capable of causing extinctions and climate perturbations. In most of the world ocean, methane, $CH_4$, continuously enters the water column from the seafloor, dissolves in seawater, and is oxidized by microbes. In some oceanic regions prone to stagnation and anoxia, methane may escape oxidation, and accumulate in the water column for a very long time, until saturation is reached. Since the saturation concentration increases with depth (due to Henry's law), a water column saturated with dissolved gas is in a metastable state (Ryskin 2003). A transition from this metastable state must eventually occur; the mechanism of transition is the water-column eruption, driven by the interplay of buoyancy forces and exsolution of dissolved gas (Ryskin 2003). A similar process is responsible for the most violent, explosive volcanic eruptions; these are driven by exsolution of water vapor dissolved in the liquid magma (Gilbert and Sparks 1998).

Extinctions are among the most important effects of the methane-driven oceanic eruptions. The eruption brings to the surface deep anoxic waters that cause extinctions in the marine realm. Terrestrial extinctions are caused by the eruption-triggered mega-tsunamis, by the explosions and conflagrations that follow the massive release of methane, and by the ensuing climate perturbations. In a large eruption, combustion and explosion of the released methane would liberate energy equivalent to $10^8$ Mt of TNT; this is greater than the world's stockpile of nuclear weapons (implicated in the nuclear-winter scenario) by a factor ~$10^4$ (Ryskin 2003).

In addition to the supporting evidence discussed in Ryskin (2003), this scenario is also in accord with other extinctions data: the preferential survival among vertebrates of the burrowing, swimming, and diving species (Sheehan and Fastovsky 1992, Retallack et al. 2003, Robertson et al. 2004), and the evidence for massive combustion of hydrocarbons (Cisowski 1990, Belcher et al. 2009).

Methane-driven oceanic eruption may produce a short-lived ocean flow of sufficient intensity to act as a hydromagnetic dynamo. Ocean water has a substantial electrical conductivity, $\sigma \approx 3.2$ S/m. One important velocity scale is the speed of propagation of tsunamis, ~200 m/s. Another is the maximum vertical velocity within the erupting fluid column, ~100 m/s. For the ocean depth $H$ of a few thousand meters, the vertical travel time is then ~1 min, whereas the time needed for a tsunami to cross the ocean is ~1 day. These timescales should be compared with the timescales of resistive decay, or magnetic diffusion. Magnetic diffusivity $\eta \equiv (\mu_0 \sigma)^{-1}$, where $\mu_0$ is a constant of SI. For ocean water $\eta \approx 0.25$ km$^2$/s, so that the timescale on which magnetic diffusion penetrates through the ocean depth is $H^2/\eta$ ~1 min. More important for dynamo action is the global diffusion timescale. For the ocean viewed as a thin spherical shell of thickness $H$ and radius $R$ (the Earth's radius), this is $RH/\eta$ ~1 day (Callarotti and Schmidt 1983). Comparison of the timescales shows that dynamo action cannot be ruled out. Below I assume that it does occur (though not necessarily in every eruption), and explore the consequences. This assumption is the only hypothetical element in the present scenario.

**2.4 Dynamo field and the lithosphere**
Direct numerical simulations show that turbulent flow of a conducting fluid can generate a large-scale magnetic field via the inverse-cascade mechanism (Brandenburg 2001). The time necessary to build up the large-scale field is comparable to the longest magnetic diffusion timescale; for the ocean it should be measured in days. The simulations also show that the overall evolution of the large-scale field is well described by the $\alpha^2$ model of the mean-field dynamo theory. This model was used by Schubert and Zhang (2001) to calculate the magnetic field generated by turbulent flow in a spherical fluid shell, with either conducting or non-conducting material occupying the interior of the shell.

As a rough approximation, the ocean during a methane-driven eruption can be viewed as such a shell. Then the interior of the shell contains the Earth's highly conducting metallic core and the essentially non-conducting mantle and crust. If the interior were entirely non-conducting, the generated magnetic field would be approximately uniform in it (Fig. 2c of Schubert and Zhang 2001). The highly conducting core changes the picture, but only slightly: the short-lived magnetic field generated by the ocean flow during the eruption ("the dynamo field") cannot enter it, except for a skin depth. The magnetic diffusivity in the core is $\eta$ ~1 m$^2$/s (Roberts and Glatzmaier 2000); the penetration depth of a dynamo field with a lifetime, say, $t = 10$ days is $(\eta t)^{1/2}$ ~1 km, whereas the radius of the core is ~3500 km. Thus, throughout its lifetime the dynamo field avoids the Earth's core.

Due to the Coriolis force dominance in the large-scale ocean flow, the direction of the dynamo field ought to be roughly aligned with the Earth's axis of rotation, with a randomly chosen polarity. (The axis of rotation provides a preferred direction, but no preferred polarity.) The magnitude of the dynamo field can be estimated by assuming that it stops growing when its back action on the flow – the Lorentz force – becomes comparable to the Coriolis force (Elsasser 1946). The ratio of the two forces is characterized by the Elsasser number $\sigma B^2/\rho\Omega$, where $B$ is the characteristic field value, $\rho$ is the density of the fluid, and $\Omega$ is the angular velocity of the Earth's rotation. The order-of-magnitude estimate of the dynamo field (given originally by Elsasser (1946) for the Earth's core model) is then $(\rho\Omega/\sigma)^{1/2}$; with $\rho$ and $\sigma$ of the ocean water this yields ~0.1 T. The dynamo field is thus greater than the geomagnetic field observed today by a factor ~$10^3$.

After the dynamo field disappears, there remains the weak field due to the lithosphere, which was magnetized by the dynamo field. To a first approximation, the lithosphere can be viewed as a spherical shell with uniform magnetic properties. The dynamo field avoids the Earth's core, but this makes it only slightly non-uniform within the lithosphere. Thus, to a first approximation, after the dynamo field disappears, a uniformly magnetized lithosphere is left. In the present scenario, the field produced by this lithosphere is the geomagnetic field observed today, and in the intervals between the dynamo-producing oceanic eruptions in the past.

## 2.5 Geomagnetic and geological implications

The earliest evidence of the geomagnetic field dates to 3.5 Gyr ago (Tarduno et al. 2010). The ocean was in existence by 3.7 Gyr ago (Fedo et al. 2001; Moorbath 2009). Methane may have been entering the ocean water column since very early times; it certainly was present in the seafloor sediments by 3.5 Gyr ago (Ueno et al. 2006), though its origin – microbial or abiotic – is a matter of dispute (Kutcherov and Krayushkin 2010).

The magnetized lithosphere being the source of the currently observed main geomagnetic field explains why the field is so stable; such stability would not be expected if the field were produced by a currently operating hydromagnetic dynamo (Zhang and Gubbins 2000). The present scenario also explains the dominance of the axial dipole, and the observed correlations between the gravitational anomalies and the long-wavelength geomagnetic anomalies. The sources of the anomalies appear to lie at the base of the crust and in the upper mantle (Pochtarev et al. 1997; Blakely et al. 2005).

The polarity of each dynamo field being arbitrary, the geomagnetic field should have reversed its direction many times throughout the Earth's history. The lifetime of the dynamo field and the duration of the polarity reversal are likely measured in days. These durations cannot be inferred on the basis of sedimentary record because during the eruption, and for some time after it, sedimentation rates can be much higher than average.

Only volcanic lava flows that erupted before or during a polarity transition, and cooled below the Curie temperature as the transition was occurring, may provide reliable information about the speed of the transition. Such lava flows have been found in Oregon, North America, in Afar, Africa, and in the Canary Islands. All show "astonishingly rapid field change" (Coe et al. 1995). In central Afar, "one lava flow has recorded both of the antipodal transitional components", and so the duration of the transition could be estimated knowing the lava's cooling rate (Acton et al. 2000). The results show that

> "directional changes during a reversal can be astonishingly fast, possibly occurring as a nearly instantaneous jump from one inclined dipolar state to another in the opposite hemisphere. … For the Site ET040 lava flow, which is ~ 4.5 m thick, to have recorded nearly antipodal transitional components above and below ~ 500°C would indicate that the jump from a northern-hemisphere transitional state to a southern-hemisphere one occurred in less than a few weeks. … Reheating and partial remagnetization by the overlying flow cannot explain either of the transitional directions because both differ significantly from that of the reversely magnetized overlying flow." (Acton et al. 2000).

In the Canary Islands, if the "peculiar" magnetization directions observed in several lava flows "could have a geomagnetic origin" (Valet et al. 1998), these flows

> "would have recorded an almost complete field reversal….Given the time taken by a 3-4 m thick lava flow to cool, this scenario would imply an extremely short duration for the reversal process." In particular, "a 160° angular deviation of the field direction would have been recorded" while the lava flow was cooling from 580°C to 500°C, a time interval that "cannot exceed …10 days" (Valet et al. 1998).

Having considered this interpretation of the data, Valet et al. (1998) rejected it because "the hypothesis of such an extremely rapid field reversal is very unlikely, if not impossible". They presented arguments in favor of remagnetization by the overlying flow, but made the following remark: "One may wonder why atypical rock magnetic properties would prevail only within flows associated with reversals."

The data of Coe et al. (1995), Acton et al. (2000), and Valet et al. (1998) cannot be reconciled with the Earth's core being the source of the geomagnetic field: the diffusion timescale of the core, ~10 kyr, would then determine the duration of reversal. In the present scenario, the Earth's core is screened from the geomagnetic field: By Runcorn's theorem, a spherical shell (the lithosphere) magnetized by a field whose sources were outside of it (the dynamo field), does not produce any field in its interior (Runcorn 1975). The core is not screened from the dynamo field, but the dynamo field is short-lived and does not enter it.

Given that methane-driven oceanic eruptions cause extinctions, polarity reversals should coincide with biostratigraphic boundaries. Not every oceanic eruption will produce a dynamo field (depending on the paleogeography, the magnitude of the eruption, etc.); even when it does, not always the resulting lithospheric field will have a polarity different from the previous one. Thus, not every biostratigraphic boundary will be marked by a geomagnetic reversal. Nor will every reversal coincide with a biostratigraphic boundary: some oceanic eruptions may result in a polarity reversal or excursion, but only a regional extinction, or none at all. Nevertheless, significant correlation between extinctions and reversals should be observable in the geological record. In addition, biostratigraphic boundaries and polarity reversals should coincide with sequence boundaries because oceanic eruptions produce mega-tsunamis.

Conflagrations and explosions that follow methane-driven oceanic eruptions produce metallic or glassy microspherules or microtektites (Cisowski 1990; Ryskin 2003); thus geomagnetic polarity reversals ought to be correlated with microtektite horizons. Such correlations are indeed observed. The Permian-Triassic boundary coincides with a polarity reversal (Ward et al. 2005) and is marked by high concentrations of microspherules (Jin et al. 2000). Preisinger et al. (2002) found magnesioferrite spinels at all four polarity reversals that they studied. Haines et al. (2004) describe microtektite-bearing deposits that coincide with the last reversal, 0.78 Myr ago. They note "abundant organic debris, including whole tree trunks", and envision rapid deposition by high-energy floods, exactly as suggested by the present scenario.

**2.6 Conclusion**

The only hypothetical element in the present scenario is the methane-driven oceanic eruption capable of dynamo action; the rest follows inescapably. Note that the present scenario is in accord with all the available geomagnetic, paleomagnetic, and geological data. In fact, the present scenario will also be in accord with most of the conventional interpretation of these data, provided one abandons the assumption of constant sedimentation rate.

Nevertheless, it is far too early to declare the problem solved. Alternative scenarios may be possible, and the present scenario may eventually be found deficient on one or more counts. A theory intended to explain phenomena in the realm of a historical science can be neither falsified nor confirmed (because experiment is impossible), and must always remain under scrutiny. But this does not mean that an attempt to build such a theory is not worth while.

Fig. 1 The geological time scale. (In this version, the smallest unit of the time scale is called "age", normally, the term "stage" is used instead; see Gradstein et al. 2004, p. 20.) Note the quasi-periodicity of the geological record over the last 542 Myr, on timescales close, in the order of magnitude, to 1 Myr. (The Precambrian part of the time scale, subdivided formally by absolute age, is not useful for present purposes.) For example, the intervals between geomagnetic polarity reversals typically range from 0.2 to 2 Myr; the intervals between geological stage boundaries - from 0.8 Myr to a few million years, etc. Exceptions do exist, e.g., there were no polarity reversals between 84 and 124 Myr ago. The origin of this quasi-periodicity, and the nature of the abrupt global events that define the boundaries of the geological time scale, are the focus of the present paper.

The original is at http://www.geosociety.org/science/timescale/

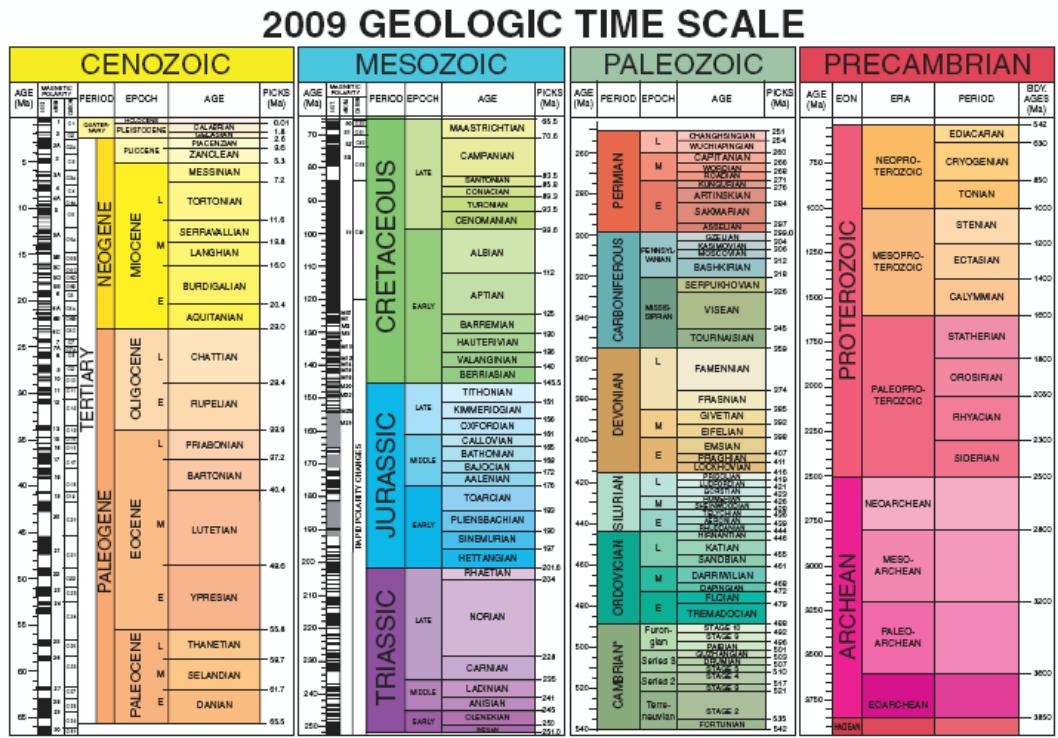

Fig. 2  The sea-level curves of sequence stratigraphy, reproduced from Haq and Schutter (2008). The original saw-tooth sea-level curve (Vail et al. 1977a, 1977b) has been renamed "onlap".  (It proved impossible to explain the abrupt sea-level falls of the original saw-tooth curve.)  The smooth version (called "sea-level changes" in the figure) is justified by invoking factors invisible in the seismic record, such as tectonic subsidence (Jervey 1988; Haq et al. 1987). Now consider the following thought experiment: If the assumption of constant sedimentation rate is dropped, the time axis can be locally stretched or compressed at will, with the sea-level curve being deformed as a result. Imagine now the sea-level curve transformed into a comb-like series of instantaneous peaks, coinciding in time with the abrupt sea-level falls of the original saw-tooth curve. In other words, could an entire cycle of sea-level rise and fall be geologically instantaneous? The answer proposed here is in the affirmative. See Section 1.7 for details.

The original is at http://www.sciencemag.org/cgi/content/full/322/5898/64 , Fig. 2

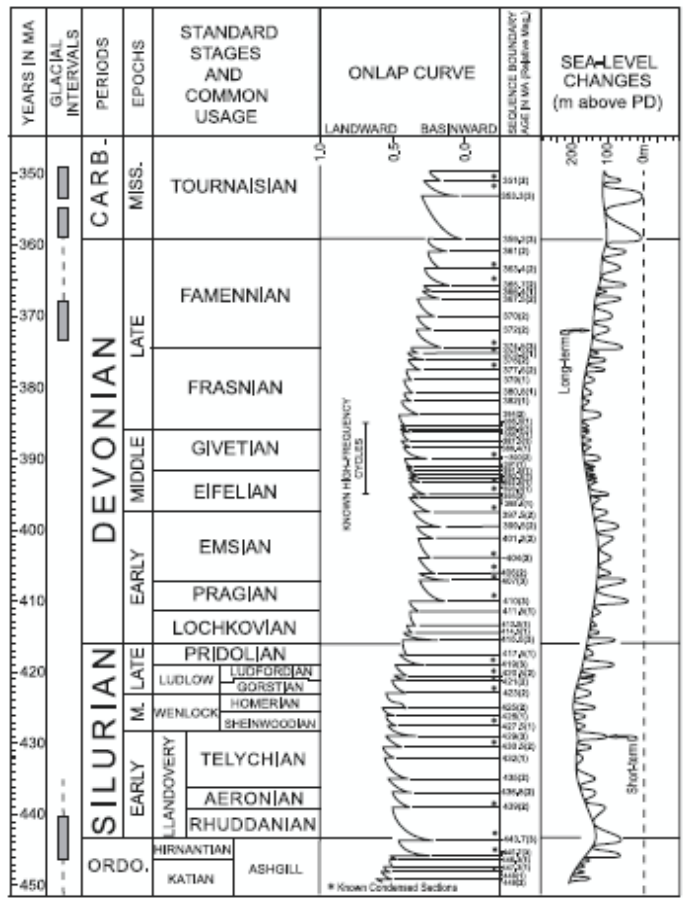

Fig. 3 An example of unconformity. An unconformity is a buried erosional surface separating strata of different ages; it indicates that sediment deposition was not continuous. Typically, the older sediments were exposed to erosion for a period of time before deposition of the younger ones, but the term is used to describe any break in the sedimentary record. In an angular unconformity, such as the one shown here, younger strata of sedimentary rock rest upon the eroded surface of tilted or folded older rocks. Not all unconformities are angular; often, the younger and the older strata are essentially parallel. Unconformities are clearly visible in the seismic record; they define sequence boundaries in sequence stratigraphy. The mapping $t(z)$ of the sedimentary column to the time axis possesses multiple discontinuities, corresponding to the periods of "lost time"; these discontinuities are marked in the sedimentary record by unconformities.

(Image © Copyright Patrick Mackie and licensed under Creative Commons License; see http://www.geograph.org.uk/photo/107618 )

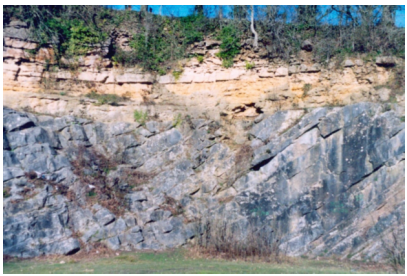